\documentclass[aps,prd,floats,floatfix,amssymb,twocolumn,superscriptaddress,nofootinbib]{revtex4-2}

\usepackage{physics}
\usepackage{amsmath}
\usepackage{amsfonts,amssymb}
\usepackage{graphicx,graphics}
\usepackage{hyperref}
\usepackage{ulem}
\usepackage{color}
\usepackage{mathrsfs}
\usepackage{bbold}
\usepackage{multirow}

\newcommand{\normalorder}[1]{\mathopen{:}#1\mathclose{:}}

\hypersetup{
  colorlinks=true,
  citecolor=blue,       
  linkcolor=blue,      
  urlcolor=cyan       
}

\makeatletter
\renewcommand{\tagform@}[1]{(\textcolor{black}{#1})}
\makeatother

\begin{document}

\title{
Tachyonic modes as a resonant system in weakly curved stellar spacetimes
}

\author{Bruno S. Felipe}
\email[]{felipe.bruno@ufabc.edu.br}
\affiliation{Centro de Matemática, Computação e Cognição, Universidade Federal do ABC (UFABC), 09210-170 Santo André, São Paulo, Brazil}

\author{Maur\'icio Richartz}
\email[]{mauricio.richartz@ufabc.edu.br}
\affiliation{Centro de Matemática, Computação e Cognição, Universidade Federal do ABC (UFABC), 09210-170 Santo André, São Paulo, Brazil}%

\date{\today}

\begin{abstract}
A real scalar field nonminimally coupled to the curvature of an astrophysical object can develop an effective potential that supports, alongside stable oscillatory solutions, a set of tachyonic modes with purely imaginary frequencies. Focusing on constant-density Newtonian stars, we show that the tachyonic sector behaves as a collection of decoupled inverted harmonic oscillators, whose quantization is naturally addressed within the rigged Hilbert space formalism. At the quantum level, this sector is described by resonant states, with mean lifetimes determined by the associated imaginary frequencies. To probe the physical implications of this framework, we compute the transition probability of an Unruh-DeWitt detector in a circular orbit. In the presence of tachyonic modes, the detector response acquires an additional finite Lorentzian profile that modifies the standard circular Unruh-like background.
\end{abstract}

\maketitle

\section{Introduction}

Quantum field theory in curved spacetime (QFTCS) stands as a remarkably successful framework for exploring quantum phenomena in gravitational backgrounds where curvature effects are significant, yet quantum-gravity corrections remain negligible~\cite{birrell1982,fulling1989,wald1994}. In stationary spacetimes, quantization can be addressed through the definition of a well-posed vacuum state via mode decomposition along a timelike Killing vector field. This vacuum notion breaks down, however, when gravity supports imaginary-frequency solutions, rendering the classical solution unstable and the conventional splitting into positive and negative-frequency modes meaningless. As a consequence, the field quantization becomes incomplete.

In astrophysical settings, such imaginary frequencies are the fundamental trigger for the spontaneous scalarization of compact objects~\cite{andreou2019,doneva2024}. During this phase transition, the field develops tachyonic excitations, characterized by an effective ``wrong-sign'' squared mass that drives the exponential growth of classical field fluctuations. A notable example occurs for nonminimally coupled scalar fields in relativistic stars, where the curvature coupling generates a tachyonic spectrum alongside the conventional oscillatory modes~\cite{lima2010,pani2011,landulfo2012,lima2013}. These modes are not optional; they belong to the complete space of solutions and must be incorporated in any consistent field expansion. The fundamental challenge, however, lies in their spectral structure, which resembles a collection of inverted harmonic oscillators whose non-square-integrable nature cannot be accommodated within the standard Hilbert space framework.

From this perspective, the quantization of tachyonic sectors requires an extension of the standard algebraic framework capable of accommodating non-square-integrable states. For unstable systems whose dynamics are governed by purely imaginary frequencies, Refs.~\cite{felipe2024,felipe2025udw} argue that the rigged Hilbert space (RHS) formalism provides the appropriate mathematical setting, naturally incorporating such modes at the quantum level. Although these works established the use of the RHS to describe bound states in non-globally hyperbolic backgrounds, its application to tachyonic instabilities induced by astrophysical objects has not yet been investigated. Addressing this gap is the primary motivation of the present work. 

In this work, we address this problem within an analytically tractable model: a real scalar field nonminimally coupled to a constant-density star in the weak-field regime ($GM/R_s \ll c^2$). Here, $G$ is Newton's constant, $c$ is the speed of light, $M$ is the mass of the star, and $R_s$ is the radius of the star. In this limit, the geometry is well described by the Newtonian approximation of General Relativity. As we will see, despite the weakness of the gravitational field, the stellar matter distribution generates an effective potential capable of supporting a discrete spectrum of unstable modes. To characterize this classical spectrum, we solve the radial field equations inside and outside the star and impose the matching conditions at the stellar surface that determine the emergence of purely imaginary frequencies.  

Once the tachyonic sector is isolated, we implement the RHS framework to identify the appropriate vacuum states and to construct a consistent Fock representation in direct analogy with canonical quantization in QFTCS. Within this approach, the tachyonic excitations acquire the interpretation of resonant states, leading to a quantum description in which the otherwise unbounded classical instability is replaced by a bounded semigroup evolution. We then explore the physical implications of this quantization by investigating the response of an Unruh-DeWitt detector in circular motion around the star. We show that the interaction between the detector and the tachyonic sector imprints a distinct resonant signature on the transition probability, modifying the standard circular Unruh-like response and providing an operational probe of the underlying vacuum structure.

This paper is organized as follows. In Sec.~\ref{sec:background} we introduce the weak-field stellar background employed throughout this work. In Sec.~\ref{sec:field-mode-solution} we solve the field equations and separate the stable and unstable sectors of the spectrum. Accordingly, Sec.~\ref{sec:quantization} constitutes the core of this work, where we develop the RHS quantization of the tachyonic modes and construct the corresponding Fock representation. In Sec.~\ref{sec:detector} we probe our results using an Unruh-DeWitt detector. Finally, Sec.~\ref{sec:conclusion} summarizes our main results and discusses future directions in which the present framework may be further explored.

\section{Background Geometry}\label{sec:background}

We consider a static, spherically symmetric object in the weak-field regime of General Relativity. Throughout this work, we keep Newton's constant $G$ explicit and adopt geometric units where $c=\hbar=1$. In isotropic coordinates $(t,r,\theta,\varphi)$, the background metric $(\mathcal{M},g)$ can be written as
\begin{equation}\label{metric-newtonian}
  ds^2 = -(1+2\Phi)dt^2 + (1-2\Phi)(dr^2+r^2d\Omega^2),
\end{equation}
where $d\Omega^2=d\theta^2+\sin^2\theta d\varphi^2$ is the standard line element of the unit two-sphere, and $\Phi=\Phi(r)$ is the Newtonian gravitational potential satisfying $|\Phi|\ll 1$ everywhere.

As an analytically tractable toy model, we consider a star composed of an incompressible fluid with uniform mass density $\rho_0$. The stress-energy tensor is thus dominated by its rest-mass density, $T_{00}\approx \rho_0$, and the background geometry is controlled to leading order by the Poisson equation $\nabla^2 \Phi = 4\pi G\rho_0$, subject to regularity at the origin and asymptotic flatness at spatial infinity.

Integrating the Poisson equation yields a piecewise-defined Newtonian potential. In the stellar interior ($r\leq R_s$), one finds
\begin{equation}
  \Phi_{\text{in}}(r)=\frac{GM}{2R_s^3}(r^2-3R_s^2),
\end{equation}
where $M=\frac{4}{3}\pi R_s^3\rho_0$ is the total mass of the star. In the exterior region ($r>R_s$), the solution matches the usual Keplerian potential
\begin{equation}
  \Phi_{\text{out}}(r)=-\frac{GM}{r}.
\end{equation}
Both $\Phi$ and its radial derivative $\Phi^{\prime}$ are continuous across the stellar surface $r=R_s$.

From Einstein's equations, the Ricci scalar $\mathcal{R}$ generated by this source is piecewise constant, namely,
\begin{equation}\label{curvature-bg-eq}
    \mathcal{R}(r) =
  \begin{cases}
    8\pi G\rho_0 = \dfrac{6GM}{R_s^3}, & r \leq R_s, \\
    0, & r > R_s.
  \end{cases}
\end{equation}
To fix scales, for a typical low-compactness astrophysical object, where $GM/R_s\sim10^{-3}$ to $10^{-6}$, the linearized metric \eqref{metric-newtonian} remains an excellent approximation throughout the entire spacetime~\cite{will2014}. Despite the small curvature, a nonminimal coupling between a scalar field and the spacetime curvature can still give rise to nontrivial effects. As we will see, sufficiently negative coupling strengths make the resulting potential well deep enough to support unstable modes.

\section{Field mode solutions}\label{sec:field-mode-solution}

Let $\phi:\mathcal{M}\to\mathbb{R}$ be a real massless scalar field governed by the Lagrangian density
\begin{equation}\label{lagrangian-eq}
  \mathcal{L}[\phi]=-\frac{1}{2}(g^{\mu\nu}\partial_{\mu}\phi\partial_{\nu}\phi+\xi \mathcal{R}\phi^2),
\end{equation}
where $\xi\in\mathbb{R}$ is the coupling constant between $\phi$ and the spacetime curvature dictated by the Ricci scalar \eqref{curvature-bg-eq}. The corresponding Euler-Lagrange equation yields the Klein-Gordon equation
\begin{equation}\label{KG-eq}
  \Box \phi - \xi \mathcal{R} \phi = 0.
\end{equation}
If $\xi=0$, we recover the usual minimally coupled scalar field, while for $\xi=1/6$ the Klein-Gordon equation and the Lagrangian density are invariant under conformal transformations.

The spherical symmetry allows us to separate variables according to
\begin{equation}\label{u-modes1}
  u_{\omega \ell m}(t,r,\theta,\varphi)=e^{-i\omega t}Y_{\ell m}(\theta,\varphi)\frac{\psi(r)}{r},
\end{equation}
where $\omega$ is the frequency and $Y_{\ell m}$ are the spherical harmonics, with $\ell$ and $m$ integers such that $|m|\leq \ell$. Substituting this ansatz into Eq.~\eqref{KG-eq}, we obtain the radial equation 
\begin{equation}\label{radial-eq}
  -\frac{d^2\psi}{dr^2} + \left[\xi \mathcal{R} + \frac{\ell(\ell+1)}{r^2} \right]\psi = \lambda \psi,\quad \lambda\equiv\omega^2,
\end{equation}
where subleading metric corrections of order $\mathcal{O}(\Phi\omega^2)$ and $\mathcal{O}(\Phi/r^2)$ have been neglected within the Newtonian approximation.

Equation~\eqref{radial-eq} defines a self-adjoint Sturm-Liouville problem with eigenvalue $\lambda\in\mathbb{R}$. For $\xi>0$, the effective potential is always repulsive, so that only scattering states exist. In contrast, for $\xi<0$, the curvature term acts as a negative square mass yielding an attractive potential inside the star. If the resulting potential well is sufficiently deep, it supports bound states with negative eigenvalues $\lambda<0$, corresponding to purely imaginary frequencies such that $\omega^2 < 0$. Consequently, the corresponding classical modes \eqref{u-modes1} present an exponential growth in time that diverges as $|t|\to\infty$. Due to the completeness of the Sturm-Liouville spectrum, these unstable solutions cannot be discarded and must be integrated into the quantum field decomposition. Since our goal is to discuss these tachyonic mode solutions, in what follows we assume $\xi<0$.

\subsection{Oscillatory field sector}
We first address the oscillatory sector with $\lambda=\omega^2>0$. In the interior, $r\leq R_s$, regularity at the origin selects the spherical Bessel function of the first kind
\begin{equation}
  r^{-1}\psi_{\text{in}}(r) = A  j_\ell(k r),
\end{equation}
as the solution of Eq.~\eqref{radial-eq}. Here, $A$ is a normalization constant, and the interior wave number is
\begin{equation}
  k^2 = \omega^2 - \xi \mathcal{R} = \omega^2 + 8\pi G |\xi| \rho_0,
\end{equation}
where in the last expression we used $\xi=-|\xi|$.

In the exterior region, $r>R_s$, the Ricci scalar vanishes, and Eq.~\eqref{radial-eq} reduces to the spherical Bessel equation. For scattering states, it is convenient to parameterize the solution in terms of a phase shift $\delta_{\ell}$:
\begin{equation}
  r^{-1}\psi_{\text{out}}(r)=\frac{B}{2i} \left[e^{-i\delta_\ell} h^{(1)}_{\ell}(\omega r)-e^{i \delta_{\ell}}h^{(2)}_{\ell}(\omega r)\right],
\end{equation}
where $h^{(1)}_{\ell},\,h^{(2)}_{\ell}$ are respectively the spherical Hankel functions defined as~\cite{olver2010}
\begin{equation}
  \begin{aligned}
    h_\ell^{(1)}&=j_\ell+iy_{\ell}\\
    h_{\ell}^{(2)}&=j_\ell-iy_{\ell},
  \end{aligned}
\end{equation}
so that they represent, respectively, outgoing and ingoing spherical waves at spatial infinity. Here, $y_{\ell}$ is the spherical Bessel function of the second kind.

Since the effective potential is finite at the stellar surface, the radial function and its first derivative are continuous at $r=R_s$:
\begin{equation}\label{matching-eq}
  \begin{aligned}
    \psi_{\text{in}}(R_s) &= \psi_{\text{out}}(R_s), \\
    \psi^{\prime}_{\text{in}}(R_s) &= \psi^{\prime}_{\text{out}}(R_s),
  \end{aligned}
\end{equation}
These conditions determine the ratio $A=A_{\omega\ell}=\psi_{\text{out}}(R_s)/\psi_{\text{in}}(R_s)$ and lead to the matching condition for the phase shift,
\begin{equation}
  \tan \delta_\ell =
  \frac{\omega\, j_\ell(k R_s)\, j_{\ell}^{\prime}(\omega R_s)-k\, j_{\ell}^{\prime}(k R_s)\, j_\ell(\omega R_s)
  }{\omega\, j_\ell(k R_s)\, y_{\ell}^{\prime}(\omega R_s)-k\, j_{\ell}^{\prime}(k R_s)\, y_\ell(\omega R_s)}.
\end{equation}

Now, we define the Klein-Gordon inner product between any two solutions $u$ and $v$ of Eq.~\eqref{KG-eq} as
\begin{equation}
  (u,v)=-i\int_{\Sigma_t}\left(u\nabla_{\mu}v^{\ast}-v^{\ast}\nabla_{\mu}u\right)n^{\mu}d\Sigma ,
\end{equation}
where $d\Sigma$ is the proper volume element on the spacelike hypersurface $\Sigma_t$ and $n^{\mu}$ is its future-directed unit normal. In the present static background, we have $n^{\mu}=\delta_0^{\mu}$, and the inner product between any two modes of the form~\eqref{u-modes1} is
\begin{equation}\label{norm-eq1}
  \begin{aligned}
    (u_{\omega \ell m},u_{\omega^{\prime}\ell^{\prime}m^{\prime}})&=(\omega+\omega^{\prime})\int d^3x\, u_{\omega \ell m}(x)u^{\ast}_{\omega^{\prime}\ell^{\prime}m^{\prime}}(x)\\
    &=(\omega+\omega^{\prime})\delta_{\ell \ell^{\prime}}\delta_{m m^{\prime}}\int_{\mathbb{R}_{+}} dr \,\psi_{\omega \ell }(r)\psi^{\ast}_{\omega^{\prime}\ell^{\prime}}(r),
  \end{aligned}
\end{equation}
where we used $\int d\Omega \,Y_{\ell m}Y^{\ast}_{\ell^{\prime}m^{\prime}}=\delta_{\ell \ell^{\prime}}\delta_{m m^{\prime}}$. To evaluate the radial normalization, it is convenient to use the asymptotic form of the spherical Hankel functions
\begin{equation}
  \psi_{\text{out}}\sim \frac{B}{\omega}\sin\left(\omega r-\frac{\pi\ell }{2}+\delta_{\ell}\right),\quad r\to \infty.
\end{equation}

The normalization \eqref{norm-eq1} enforces $B=\sqrt{\omega/\pi}$, ensuring that the oscillatory modes satisfy the orthonormality relations
\begin{equation}
    \begin{aligned}
        (u_{\omega \ell m},u_{\omega^{\prime}\ell^{\prime}m^{\prime}})&=\delta(\omega-\omega^{\prime})\delta_{\ell \ell^{\prime}}\delta_{m m^{\prime}},\\
        (u_{\omega \ell m}^{\ast},u_{\omega^{\prime}\ell^{\prime}m^{\prime}}^{\ast})&=-\delta(\omega-\omega^{\prime})\delta_{\ell \ell^{\prime}}\delta_{m m^{\prime}}.
    \end{aligned}
\end{equation}
In this way, the stable oscillatory field sector can be expanded as
\begin{equation}\label{phi-a-eq}
  \phi_a(x)=\sum_{\ell m}\int d\omega \left[a_{\omega \ell m}u_{\omega \ell m}(x)+a^{\dagger}_{\omega \ell m}u_{\omega \ell m}^{\ast}(x)\right]
\end{equation}
where $a_{\omega \ell m}=(u_{\omega \ell m},\phi_a)$ and $a^{\dagger}_{\omega \ell m}=-(u_{\omega \ell m}^{\ast},\phi_a)$ are expansion coefficients. 

Quantization then proceeds by promoting $\phi_a$ and its conjugate momentum to operator-valued distributions acting upon the corresponding Fock space. This is equivalent to imposing the commutation relations
\begin{equation}
  \begin{aligned}
      [a_{\omega \ell m},a^{\dagger}_{\omega^{\prime}\ell^{\prime}m^{\prime}}]
  &=\delta(\omega-\omega^{\prime})\delta_{\ell\ell^{\prime}}\delta_{mm^{\prime}},\\
  [a_{\omega \ell m},a_{\omega^{\prime}\ell^{\prime}m^{\prime}}]&=[a^{\dagger}_{\omega \ell m}, a^{\dagger}_{\omega^{\prime}\ell^{\prime}m^{\prime}}]=0.
  \end{aligned}
\end{equation}
This algebraic structure unambiguously identifies $a_{\omega \ell m}$ and $a_{\omega \ell m}^{\dagger}$ as annihilation and creation operators, respectively. Accordingly, we define the vacuum state $|0\rangle$ as
\begin{equation}\label{vacuum-eq}
  a_{\omega \ell m}|0\rangle=0,\quad \forall \omega, \ell, m.
\end{equation}

The corresponding excited states are then constructed by successive applications of the creation operator on the vacuum, leading to the $n$-particle state
\begin{equation}\label{excited-free-states}
  |n_{\omega \ell m}\rangle=\frac{a_{\omega \ell m}^{\dagger}\dots a_{\omega \ell m}^{\dagger}}{\sqrt{n!}}|0\rangle.
\end{equation}
These states (including all $n$-particle excitations) span the Fock space $\mathfrak{F}_a$, which is the space of all possible oscillatory particles described by static observers. 

\subsection{Tachyonic field sector}

Completeness also requires that the sector with $\lambda=\omega^2<0$ be included in the field expansion. Writing $\omega=\pm i\Omega$, with $\Omega>0$, the interior solution of Eq.~\eqref{radial-eq} takes the form
\begin{equation}
  r^{-1}\psi_{\text{in}}(r)=C j_{\ell}(q r),
\end{equation}
where
\begin{equation}\label{q-eq}
  q^2= 8\pi G |\xi|\rho_0-\Omega^2.
\end{equation}
In the exterior region, on the other hand, the replacement $\omega^2\to-\Omega^2$ leads to
\begin{equation}\label{psi-out}
  r^{-1}\psi_{\text{out}}(r)=D k_{\ell}(\Omega r)+E i_{\ell}(\Omega r),
\end{equation}
where $i_{\ell}(z)=i^{-\ell}j_{\ell}(iz)$ and $k_{\ell}(z)=-i^{\ell}h_{\ell}^{(1)}(iz)$ are the modified spherical Bessel functions of the first and second kinds, respectively. Imposing square integrability sets $E=0$.

Applying the matching conditions~\eqref{matching-eq} determines the ratio
$C/D=j_{\ell}(qR_s)/k_{\ell}(\Omega R_s)$ and yields the following transcendental equation
for $\Omega$,
\begin{equation}\label{matching2-eq}
  x \, \frac{j_{\ell+1}(x)}{j_{\ell}(x)}
  =
  y \, \frac{k_{\ell+1}(y)}{k_{\ell}(y)},
\end{equation}
where we used the recurrence relation
$z f_{\ell}^{\prime}(z)=\ell f_{\ell}(z)-z f_{\ell+1}(z)$ for
$f_{\ell}=j_{\ell},\,k_{\ell}$~\cite{abramowitz}, and introduced the
dimensionless variables $x=qR_s$ and $y=\Omega R_s$, which satisfy
\begin{equation}
  x^2+y^2=\frac{6G|\xi|M}{R_s}.
\end{equation}

\begin{figure}[!htb]
  \centering
  \includegraphics[width=0.4\textwidth]{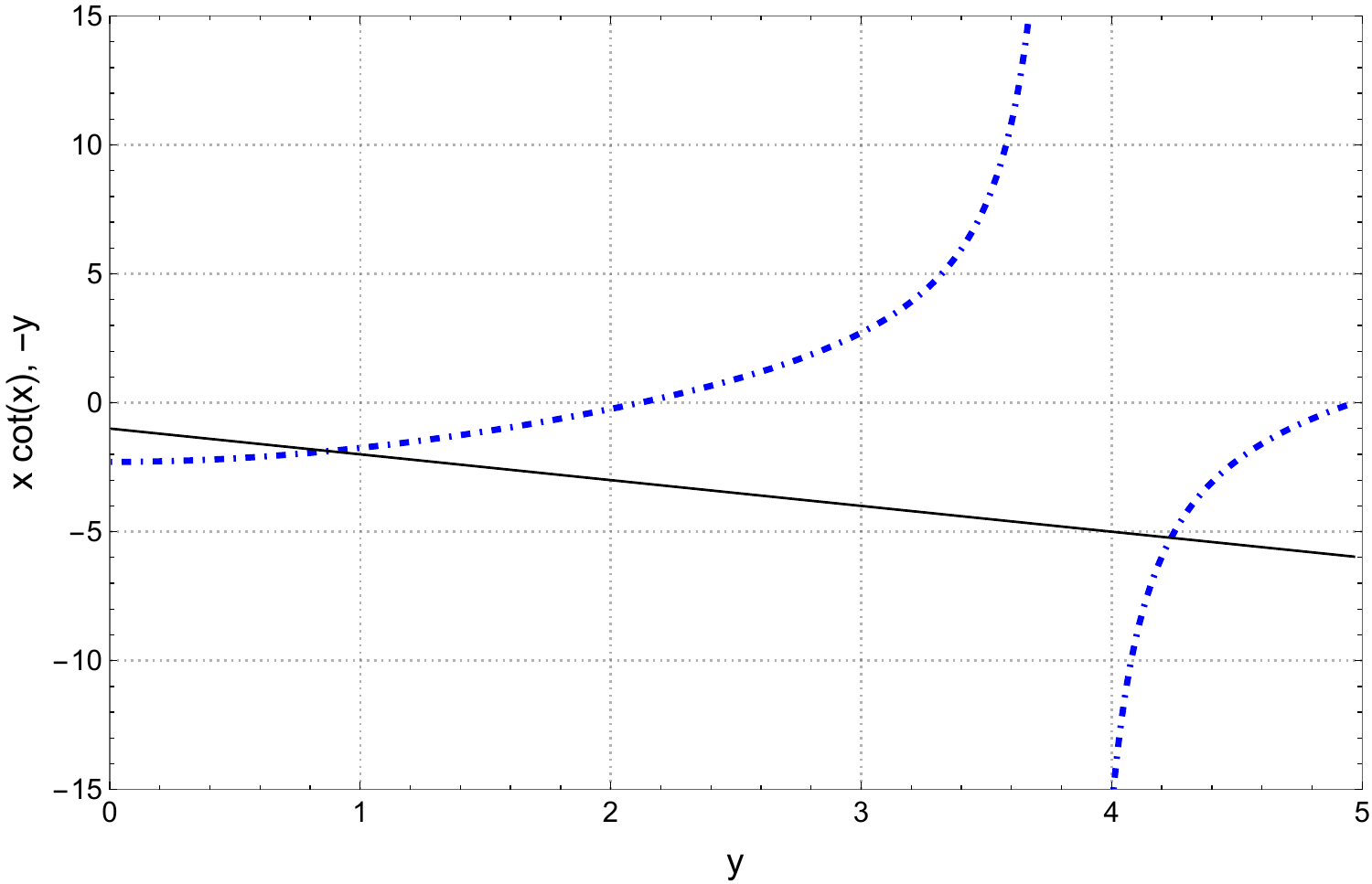}
  \caption{The roots of Eq.~\eqref{l=0-eq} for the coupling constant $\xi=10\xi_c=-10\pi^2/4$, obtained from the intersections of the solid line $-y$ with the dot-dashed curves $x\cot(x)$. The allowed frequencies in this case are $y=\Omega R_s=0.87$ and $4.23$.}
  \label{fig:omegas}
\end{figure}

In the simplest case $\ell=0$, in which the centrifugal barrier is absent, Eq.~\eqref{matching2-eq} reduces to
\begin{equation}\label{l=0-eq}
  x\cot(x)=-y.
\end{equation}
Fig.~\ref{fig:omegas} shows the graphical solution of Eq.~\eqref{l=0-eq} for a particular value of $\xi$. The intersections of the curves determine the allowed values of $y=\Omega R_s$.

At the threshold at which $\omega^2$ changes sign (i.e., $\Omega=0$), the matching condition \eqref{l=0-eq} implies $\cot(x)=0$. Using \eqref{q-eq}, this yields the critical coupling
\begin{equation}
  \xi_{c}=-\frac{\pi^2}{24\left(\frac{GM}{R_s}\right)}.
\end{equation}
Consequently, unstable (tachyonic) modes occur only when the coupling is sufficiently negative, i.e., $\xi<\xi_{c}$, a behavior qualitatively consistent with the onset of tachyonic instabilities in fully relativistic stars~\cite{pani2011}. Notably, the threshold scales inversely with the stellar compactness, meaning that weaker gravitational backgrounds require a larger nonminimal coupling magnitude to trigger the instability. 

Let $\Omega_{n\ell}$ be a solution of Eq.~\eqref{matching2-eq}. The radial function can then be normalized in $\mathrm{L}^2(\mathbb{R}_{+},dr)$ according to
\begin{equation}\label{normalization-psi}
    1=\int_0^{R_s}|\psi_{\text{in}}(r)|^2 dr+\int_{R_s}^{\infty}|\psi_{\text{out}}(r)|^2dr,
\end{equation}
which gives
\begin{equation}
    \begin{aligned}
        |C_{n\ell}|^{-2}&=\int_0^{R_s}|j_{\ell}(q_{n\ell} r)|^2 r^2dr \\
    &\quad+\left|\frac{k_{\ell}(\Omega_{n\ell} R_s)}{j_{\ell}(q_{n\ell} R_s)}\right|^2\int_{R_s}^{\infty}|k_{\ell}(\Omega_{n\ell} r)|^2 r^2 dr.
    \end{aligned}
\end{equation}

At this point, the classical unstable modes are fully characterized. Since $\omega^2=-\Omega_{n\ell}^2<0$, the time evolution of the system involves both growing and decaying real exponentials, $e^{+\Omega_{n\ell} t}$ and $e^{-\Omega_{n\ell} t}$, and hence the complete tachyonic sector must include both branches. Therefore, introducing the modes
\begin{equation}
  v^{(\pm)}_{n\ell m}(x)=\frac{e^{\mp \Omega_{n\ell} t}}{\sqrt{2\Omega_{n\ell}}}Y^{\pm}_{\ell m}(\theta,\varphi) r^{-1}\psi_{n \ell}(r),
\end{equation}
where $Y_{\ell m}^{+}=Y_{\ell m}$ and $Y^{-}_{\ell m}=Y_{\ell m}^{\ast}$, the unstable contribution can be expressed as
\begin{equation}\label{phi-b-eq}
  \phi_b(x)=\sum_{n\ell m}\left[b_{n\ell m}^{+}v_{n \ell m}^{(+)}(x)+b_{n\ell m}^{-}v_{n \ell m}^{(-)}(x)\right],
\end{equation}
where $b^{\pm}_{n\ell m}$ are expansion coefficients.

Unlike the oscillatory case, $\phi_b$ is not expanded in terms of positive- and negative-frequency modes. Instead, a direct computation shows that the corresponding unstable modes satisfy a non-standard cross-orthonormality structure under the Klein--Gordon inner product
\begin{equation}\label{normalization-v}
  \begin{aligned}
      (v^{(\pm)}_{n\ell m},v^{(\mp)}_{n \ell^{\prime}m^{\prime}})&=\mp i\,\delta_{\ell \ell^{\prime}}\delta_{m m^{\prime}},\\\quad
  (v^{(\pm)}_{n\ell m},v^{(\pm)}_{n \ell^{\prime}m^{\prime}})&=0,
  \end{aligned}
\end{equation}
and hence they are ``imaginary normalized.'' Because the expansion coefficients $b_{n\ell m}^{\pm}=\pm i(\phi_b,v^{(\mp)}_{n\ell m})$ are kinematically coupled and lack the algebraic properties of independent creation and annihilation operators, the standard vacuum notion \eqref{vacuum-eq} becomes ill-defined in this sector. Thus, the associated classical time divergence renders the canonical quantization scheme incomplete and requires a more general mathematical framework to consistently construct the full quantum kinematics.

In summary, the complete classical field configuration is formally decomposed into its stable and unstable spectral components as
\begin{equation}
  \phi(x)=\phi_a(x)+\phi_b(x).
\end{equation}
While the oscillatory contribution $\phi_a(x)$ admits a standard canonical quantization leading to the conventional Fock space $\mathfrak{F}_a$, the non-oscillatory sector $\phi_b(x)$ evades this procedure due to the exponential divergence of its modes with imaginary frequencies. This limitation of the canonical quantization procedure requires a more general quantization framework. In the following sections, we demonstrate that, by employing the RHS, one can consistently define generalized creation and annihilation operators for the tachyonic sector.

\section{Tachyonic sector quantization}\label{sec:quantization}

In the stable oscillatory sector, quantization is guided by the ordinary harmonic oscillator analogy, which naturally provides a vacuum state and a Fock representation. For the unstable expansion in Eq.~\eqref{phi-b-eq}, the relevant mechanical analog is instead the {\it inverted harmonic oscillator}, obtained formally by the replacement $\omega\to\pm i\Omega$.

To make this correspondence precise, let $I=(n,\ell,m)$ collectively denote the mode labels, define $\Omega_I\equiv\Omega_{n\ell}$, and set $b_I^{\pm}(t)=b_I^{\pm}e^{\mp \Omega_I t}$. Hereafter, the collective index $I$ is used whenever no ambiguity arises, while explicit mode labels are used in mode expansions. This allows us to write the unstable contribution \eqref{phi-b-eq} in the form
\begin{equation}\label{alternative-modes}
  \phi_b(t,{\bf x})=\sum_{I} \chi_{I}(t)
  Z_{I}(\theta,\varphi)r^{-1}\psi_{I}(r),
\end{equation}
where $Z_I=Z_{\ell m}$ are the real spherical harmonics and
\begin{equation}\label{chi-i-def}
  \chi_{I}(t):=\frac{b^+_I(t)+b_I^-(t)}{\sqrt{2\Omega_I}}.
\end{equation}
Inserting $\phi_b$ into the Lagrangian density \eqref{lagrangian-eq} and integrating over the space, we find the effective Lagrangian (see Appendix~\ref{app:effective-lagrangian})
\begin{equation}\label{lagrangian-phib}
  L[\phi_b]=\int d^3x \sqrt{-g}\, \mathcal{L}[\phi_b]
  =\sum_{I}\frac{1}{2}\left(\dot\chi_{I}^2+\Omega_I^2 \chi_I^2\right),
\end{equation}
with $\dot\chi_I\equiv\partial_t\chi_I$. The canonical dynamics is then described by conjugate variables $(\chi_I,P_I)$ with Hamiltonian
\begin{equation}\label{HI-eq}
  H_b=\sum_{I}H_I,
  \quad
  H_I=\frac{1}{2}\left(P_I^2-\Omega_I^2 \chi_I^2\right),
\end{equation}
where $P_I=\partial L/\partial\dot \chi_I=\dot \chi_I$. Precisely, each contribution $H_I$ is the Hamiltonian of a unit-mass one-dimensional inverted harmonic oscillator, and the temporal domain of the tachyonic field sector evolves in time as a collection of decoupled inverted harmonic oscillators.

Unlike the ordinary oscillatory case, the tachyonic Hamiltonian in Eq.~\eqref{HI-eq} is unbounded from below and therefore admits no normalizable ground state in the usual Hilbert space sense. From this point of view, it is clear that the Fock space generated by the standard vacuum $|0\rangle$ in Eq.~\eqref{vacuum-eq} is incompatible with the tachyonic Hamiltonian. A suitable framework for treating this sector is provided instead by the rigged Hilbert space formalism~\cite{delamadrid2007,delamadrid2005,bohm1981,gadella1983,antoniou2001}, which extends the traditional inner product space of square-integrable functions to distributional spaces that allow the description of scattering, decay, and resonant phenomena through exact solutions to eigenvalue equations.

\subsection{The rigged Hilbert space formulation}
The RHS is characterized by the Gelfand triplet of spaces
\begin{equation}\label{RHS}
  \mathrm{S}\subset \mathcal{H}\subset \mathrm{S}^{\times},
\end{equation}
where $\mathcal{H}$ is an infinite-dimensional Hilbert space, $\mathrm{S}$ is a dense subspace of $\mathcal{H}$ endowed with a nuclear topology, and $\mathrm{S}^{\times}$ is the dual space of $\mathrm{S}$, consisting of continuous linear functionals. Here, ``rigged'' refers precisely to the additional topological structure carried by the triplet: $\mathcal{H}$ is equipped with a topology $\tau_{\mathcal{H}}$, $\mathrm{S}$ is endowed with a finer topology $\tau_{\mathrm{S}}$, and the dual space $\mathrm{S}^{\times}$ inherits the corresponding weaker topology $\tau_{\mathrm{S}^{\times}}$.

As extensively discussed by A. Bohm~\cite{bohm_dollard}, Dirac's {\it braket} formalism retains its full operational validity within the RHS framework. This is achieved by interpreting the {\it ket} states as elements of the dual space $\mathrm{S}^{\times}$, and the {\it bra} states as elements of the dense test-function space $\mathrm{S}$. This strict assignment ensures that the {\it braket} operation is well-defined and mathematically convergent\footnote{Recall that in standard quantum mechanics, the Riesz-Fréchet theorem~\cite{conway1994} establishes an isomorphism between the Hilbert space and its dual $(\mathcal{H}\approx\mathcal{H}^{\times}
)$. This allows one to naturally identify bras and kets without topological ambiguities. In the RHS framework, however, treating them as elements of distinct functional spaces becomes strictly necessary.}. 

In the RHS formalism, if an operator $A$ leaves $\mathrm{S}$ invariant, then it admits a dual extension $A^{\times}$ defined by
\begin{equation}\label{duality-eq}
  \langle A\varphi \mid F\rangle
  =
  \langle \varphi \mid A^{\times}F\rangle,
  \quad
  \varphi\in\mathrm{S},\quad F\in\mathrm{S}^{\times}.
\end{equation}
Accordingly, a complex number $\lambda$ is called a generalized eigenvalue of $A$ if there exists a nonzero vector state $|F_\lambda\rangle\in\mathrm{S}^{\times}$ such that, for every $\varphi\in\mathrm{S}$,
\begin{equation}
  \langle A\varphi \mid F_\lambda\rangle
  =
  \langle \varphi \mid A^{\times}F_\lambda\rangle
  =
  \lambda\langle \varphi \mid F_\lambda\rangle.
\end{equation}

Suppressing the arbitrary test vector $\varphi$, we may write this relation in the familiar form
\begin{equation}
  A^{\times}|F_\lambda\rangle=\lambda |F_\lambda\rangle,
\end{equation}
where $\lambda$ may be either real or complex, since the generalized eigenvector belongs to the larger space $\mathrm{S}^{\times}$. In this sense, the RHS framework extends the usual Hilbert-space formulation so that it is possible to accommodate the spectral structure relevant for the inverted harmonic oscillator. With this structure in place, we can analyze the quantum inverted harmonic oscillator in two complementary representations: the energy basis and the decaying/growing basis.

\subsection{Spectral representations of the inverted harmonic oscillator}

To analyze the quantum dynamics of a single tachyonic mode governed by the Hamiltonian $H_I$ of Eq.~\eqref{HI-eq}, we introduce the canonical commutation relation
\begin{equation}\label{commutation-chi-p}
  [\chi_I,P_{J}]=i\delta_{IJ}.
\end{equation}
The generalized eigenspace associated with $H_I$ is doubly degenerate: for each $E\in\mathbb{R}$ there are two generalized eigenvectors, denoted by $|E^I_{\pm}\rangle$, satisfying
\begin{equation}
  H_I^{\times}|E^I_{\pm}\rangle=E|E^I_{\pm}\rangle.
\end{equation}

In the $\chi_I$ representation, i.e., $\psi^E_{\pm}(\chi_I)\equiv\langle \chi_I|E^I_{\pm}\rangle$, the corresponding eigenvalue equation is solved by linear combinations of parabolic cylinder functions, with the relevant basis fixed by the chosen boundary conditions. For our purposes, we take~\cite{chruscinski2003}
\begin{equation}\label{energy1}
  \psi^E_{\pm}(\chi_I)
  =
  \frac{C_{I}}{\sqrt{2\pi\Omega_I}}\,
  i^{\frac{\nu_I+1}{2}}\,
  \Gamma(\nu_I+1)
  D_{-\nu_I-1}\!\left(\mp \sqrt{-2i\Omega_I}\,\chi_I\right),
\end{equation}
where
\begin{equation}
  \nu_I=-\left(i\frac{E}{\Omega_I}+\frac{1}{2}\right),
  \quad
  C_{I}=\left(\frac{\Omega_I}{2\pi^2}\right)^{1/4}.
\end{equation}
These eigenfunctions are completed by the conjugate family $\overline{\psi^E_{\pm}}$, which satisfies the generalized eigenvalue equation $H_I^{\times}\overline{\psi^E_{\pm}}=-E\,\overline{\psi^E_{\pm}}$. 

Because the parabolic cylinder functions $D_{\nu_I}$ diverge exponentially as $|\chi_I|\to\infty$, the function in~\eqref{energy1} and its conjugate are not square integrable and therefore do not belong to the Hilbert space $\mathcal{H}=\mathrm{L}^2(\mathbb{R}_{\chi_I})$. Nevertheless, they obey orthogonality and completeness relations in the distributional sense:
\begin{equation}
  \begin{aligned}
    \int_{\mathbb{R}} \overline{\psi^E_{\pm}(\chi_I)}\,\psi^{E^{\prime}}_{\pm}(\chi_I)\,d\chi_I
    &=\delta(E-E^{\prime}),\\
    \int_{\mathbb{R}} \overline{\psi^E_{\pm}(\chi_I)}\,\psi^E_{\pm}(\chi_I^{\prime})\,dE
    &=\delta(\chi_I-\chi_I^{\prime}).
  \end{aligned}
\end{equation}
For this reason, the natural RHS associated with the $I$-th inverted oscillator is obtained by identifying the subspace $\mathrm{S}$ as the Schwartz space $\mathrm{S}(\mathbb{R}_{\chi_I})$, namely, the space of rapidly decreasing functions. Its dual, $\mathrm{S}^{\times}(\mathbb{R}_{\chi_I})$, is the space of tempered distributions, in which the generalized eigenvectors $|E^I_{\pm}\rangle$ are well defined. Accordingly, the corresponding triplet is
\begin{equation}
  \mathrm{S}(\mathbb{R}_{\chi_I})\subset \mathrm{L}^2(\mathbb{R}_{\chi_I})\subset \mathrm{S}^{\times}(\mathbb{R}_{\chi_I}).
\end{equation}

Alternatively, in analogy with the ordinary harmonic oscillator, one can construct a discrete representation by introducing the canonical ladder operators
\begin{equation}\label{b+- definition}
  b^{\pm}_I := \sqrt{\frac{\Omega_I}{2}} \left( \chi_I \mp \frac{P_I}{\Omega_I} \right).
\end{equation}
On this basis, the commutation relation \eqref{commutation-chi-p} implies that
\begin{equation}\label{commutation-bb}
  [b^{+}_I, b^{-}_J] = i\delta_{IJ}, \quad [b^{\pm}_I, b^{\pm}_J] = 0,
\end{equation}
and the Hamiltonian $H_I$ becomes
\begin{equation}
  H_I = -\frac{\Omega_I}{2}(b^{+}_Ib^{-}_I + b^{-}_Ib^{+}_I).
\end{equation}

These operators naturally admit dual extensions $(b_I^{\pm})^{\times}$ via Eq.~\eqref{duality-eq}. Their action on an arbitrary generalized state $F\in\mathrm{S}^{\times}$ is given by $(b_I^{\pm})^{\times}F=b^{\pm}_IF$ (see Ref.~\cite{shimbori2000}), where the test-function space is identified with the Schwartz space $\mathrm{S}=\mathrm{S}(\mathbb{R}_{\chi_I})$. Proceeding algebraically as in the ordinary harmonic oscillator, one defines generalized vacuum states $|0^{\pm}\rangle$ by
\begin{equation}\label{vacua+-}
  (b^{\pm}_I)^{\times}|0^{\mp}\rangle= 0.
\end{equation}
Hence, the extension of $b^+_I$ annihilates $|0^-\rangle$, while the extension of $b^-_I$ annihilates $|0^+\rangle$.

From these vacuum states, one can construct excited states in a similar way. Let these states be

\begin{equation}\label{excited-states+}
  |n^{+}_I\rangle = \frac{1}{\sqrt{n!}}[(b^{+}_I)^{\times}]^n |0^{+}\rangle
\end{equation}
and
\begin{equation}\label{excited-states-}
  |n^{-}_I\rangle=\frac{(-i)^n}{\sqrt{n!}}[(b_I^-)^{\times}]^n|0^{-}\rangle,
\end{equation}
for $n\in \mathbb{Z}^+_0$. By construction, these states are generalized eigenvectors of $H_I$ with purely imaginary eigenvalues:
\begin{equation}\label{hamiltonian2}
  H^{\times}_I|n^{\pm}_I\rangle = \pm E_{n_I} |n^{\pm}_I\rangle, \quad E_{n_I} = i\Omega_I \left( n + \frac{1}{2} \right).
\end{equation}

In the $\chi_I$ representation, i.e., $f_{n_I}^{\pm}(\chi_I)\equiv\langle \chi_I|n_I^{\pm}\rangle$, Eq.~\eqref{hamiltonian2} gives
\begin{equation}\label{fn-position}
  f_{n_I}^{\pm}(\chi_I) = N_n^{\pm} e^{\mp \frac{i\Omega_I \chi_I^2}{2}} H_n(\sqrt{\pm i\Omega_I} \chi_I),
\end{equation}
where $H_n$ are Hermite polynomials and $N_n^{\pm}$ are normalization constants. These generalized wave functions satisfy the following properties:
\begin{enumerate}
  \item {\it Conjugacy}\label{property1} --- They are related by complex conjugation
    \begin{equation}
      \overline{f_{n_I}^+(\chi_I)}=f_{n_I}^-(\chi_I).
    \end{equation}
  \item {\it Biorthogonality}\label{property2} --- They satisfy the biorthogonality relation
    \begin{equation}
      \int_{\mathbb{R}_{\chi_I}}\overline{f^{\pm}_{n_I}(\chi_I)}f_{m_I}^{\mp}(\chi_I)d\chi_I=\delta_{n_Im_I}.
    \end{equation}
  \item {\it Completeness}\label{property3} --- They satisfy the completeness relation
    \begin{equation}
      \sum_{n=0}^{\infty}\overline{f_{n_I}^{\pm}(\chi_I)}f_{n_I}^{\mp}(\chi^{\prime}_I)
      =\delta(\chi_I-\chi^{\prime}_I).
    \end{equation}
\end{enumerate}

Properties~\ref{property1}--\ref{property3} show that the two families $\{f_{n_I}^+\}$ and $\{f_{n_I}^-\}$ constitute a complete biorthogonal system. In particular, property~\ref{property2} implies that the bra associated with $|n_I^{\pm}\rangle$ is $\langle n_I^{\mp}|$, mirroring the cross-normalization of the classical modes $v_I^{(\pm)}$ in Eq.~\eqref{normalization-v}. The completeness relation, property~\ref{property3}, is understood in the distributional sense, since these generalized states do not belong to the Hilbert space $\mathscr{H}=\mathrm{L}^2(\mathbb{R}_{\chi_I})$.

To characterize the domains associated with $f_{n_I}^{+}$ and $f_{n_I}^{-}$, it is useful to introduce two triplets of spaces,
\begin{equation}\label{triplet22}
  \mathrm{S}_{+}\subset\mathscr{H}\subset\mathrm{S}_{+}^{\times} \quad \text{and}\quad \mathrm{S}_{-}\subset\mathscr{H}\subset\mathrm{S}_{-}^{\times},
\end{equation}
where $|n^{+}_I\rangle\in \mathrm{S}_{+}^{\times}$ and $|n^{-}_I\rangle\in \mathrm{S}_{-}^{\times}$, with $\mathrm{S}_{+}\cap\mathrm{S}_{-}=\{\emptyset\}$.

The spaces $\mathrm{S}_{\pm}$ were rigorously defined by Chruściński~\cite{chruscinski2003,chruscinski2004}. Analytic continuation of $\psi_{\pm}^{E}$ and $\overline{\psi_{\pm}^{E}}$ to the complex $E$-plane shows that they have simple poles at $E=\mp i\Omega_I(n+1/2)$, whose residues define the states $f_{n_I}^{\pm}$. This leads to the characterization
\begin{equation}\label{space-definition}
  \begin{aligned}
    \mathrm{S}_{-}&=\left\{\varphi \in \mathrm{S}(\mathbb{R}_{\chi_I}) \,\big| \langle\varphi|\overline{\psi_{\pm}^{E}}\rangle\in \mathrm{S}(\mathbb{R}_{E})\cap\mathscr{H}^{2}_{-}(\mathbb{R}_{E})\right\}\\
    \text{and} &\\
    \mathrm{S}_{+}&=\left\{\varphi \in \mathrm{S}(\mathbb{R}_{\chi_I}) \,\big|\langle\varphi|\psi_{\pm}^{E}\rangle\in \mathrm{S}(\mathbb{R}_{E})\cap\mathscr{H}^{2}_{+}(\mathbb{R}_{E})\right\},
  \end{aligned}
\end{equation}
where $\mathscr{H}_{+}^2(\mathscr{H}_{-}^2)$ denotes the Hardy class space~\cite{duren2000} for the upper (lower) half-plane. Specifically, $\mathrm{S}_{+}$ consists of functions that are boundary values of analytic functions in the upper half-plane of the complex $E$-plane and vanish faster than any power of $E$ along the upper semicircle. Similarly, $\mathrm{S}_{-}$ corresponds to the analogous space for functions analytic in the lower half-plane.

These results allow us to interpret $|n^{\pm}_I\rangle$ as generalized resonant states which, in close analogy with canonical quantization, are systematically suited to represent the tachyonic degrees of freedom. Consequently, we can now generalize these single-mode findings to the complete multi-mode collection of the unstable sector.

\subsection{Tachyonic field sector representation}

To accommodate arbitrary numbers of tachyonic excitations across the complete decoupled collection of inverted harmonic oscillators, we interpret the composite multi-mode states through a symmetric tensor product structure. Formally, a generalized state with $n$-excitations in a specific mode $I$ is defined via occupation numbers as $|\{n_K = n\delta_{KI}\}^{\pm}\rangle \equiv |0, \dots, n, \dots, 0\rangle^{\pm} = |0_1^{\pm}\rangle \otimes \cdots \otimes |n_I^{\pm}\rangle \otimes \cdots \otimes |0_N^{\pm}\rangle$, which we denote simply as $|n_I^{\pm}\rangle$. This state acts as a generalized eigenvector of the total dual Hamiltonian $H_b^{\times} = \sum_I H_I^{\times}$, satisfying:
\begin{equation}
  H_b^{\times}|n_I^{\pm}\rangle
  =
  \pm i\left(n\Omega_I+\frac{1}{2}\sum_K\Omega_K\right)|n_{I}^{\pm}\rangle.
\end{equation}

In this expression, the first term on the right-hand side corresponds to the purely imaginary energy contribution of the excited state labeled by $I=\{n,\ell,m\}$, while the second term represents the (possibly divergent) zero-point energy sum over the entire inverted harmonic oscillator collection. Because we are treating the scalar field as a test field on a fixed background geometry, its absolute zero-point energy does not source the background gravitational field. Consequently, regardless of whether this vacuum sum converges or diverges, we can safely remove it via normal ordering with respect to the chosen vacuum. This ensures our focus remains solely on the physically meaningful resonant excitations. In our weak-field stellar scenario ($\Phi\ll 1$), the resulting field configuration effectively mirrors a flat-space evolution subject to regularity constraints at the origin. Accordingly, the total dual Hamiltonian can be regularized via normal ordering, defined through the matrix elements
\begin{equation}
  \langle n_I^{\mp}|\normalorder{H_b^{\times}}|n_I^{\pm}\rangle
  \equiv
  \langle n_I^{\mp}|H_b^{\times}|n_I^{\pm}\rangle
  -
  \langle 0^{\mp}|H_b^{\times}|0^{\pm}\rangle,
\end{equation}
which successfully subtracts the zero-point contribution to yield
\begin{equation}
  \normalorder{H_b^{\times}}|n_I^{\pm}\rangle=\pm i n\Omega_I|n_{I}^{\pm}\rangle.
\end{equation}

To accommodate an arbitrary occupation number, one introduces the symmetric tensor product spaces
\begin{equation}
  \mathfrak{F}_n(\mathcal{H})\equiv(\bigotimes {}^n\mathcal{H})_s,
  \quad
  \mathfrak{F}_n(\mathrm{S})\equiv(\bigotimes {}^n\mathrm{S})_s,
\end{equation}
with analogous definitions for $\mathfrak{F}_n^{\times}(\mathrm{S}^{\times})$. Here, $\mathrm{S}=\mathrm{S}_{+}\cup\mathrm{S}_{-}$. The tachyonic quantum states are therefore described by the rigged Fock triplet
\begin{equation}\label{triplet-2}
  \mathfrak{F}_b(\mathrm{S})\subset\mathfrak{F}(\mathcal{H})\subset\mathfrak{F}_b^{\times}(\mathrm{S}^{\times}),
\end{equation}
where the full spaces are defined via the infinite direct sums
\begin{equation}
\mathfrak{F}_b(\mathrm{S}) = \bigoplus_{n=0}^{\infty} \mathfrak{F}_n(\mathrm{S}), \qquad
\mathfrak{F}(\mathcal{H}) = \bigoplus_{n=0}^{\infty} \mathfrak{F}_n(\mathcal{H}),
\end{equation}
and $\mathfrak{F}_b^{\times}(\mathrm{S}^{\times})$ is defined as the dual space of $\mathfrak{F}_b(\mathrm{S}).$

The coefficients $b^{\pm}_I$ in Eq.~\eqref{phi-b-eq} are then interpreted as extended ladder operators between neighboring spaces $(b^{\pm}_I)^{\times}:\mathfrak{F}_{n}^{\times} \longrightarrow \mathfrak{F}_{n \pm 1}^{\times}$, and the tachyonic field is understood as a generalized operator acting on the Fock triplet~\eqref{triplet-2}.

\subsection{Semigroups dynamics and vacuum fluctuation}

The restriction of the test-function spaces to the Hardy classes $\mathrm{S}_{\pm}$ fundamentally alters the temporal quantum dynamics, reducing the standard unitary time-evolution group to a pair of strictly directed semigroups when acting on generalized states~\cite{chruscinski2003}. Let $U_I(t) = e^{-iH_I t}$ be the single-mode time-evolution operator. For a regular test state $\varphi^{\pm} \in \mathrm{S}_{\pm}$, the duality relation with the generalized resonance states yields (see also Ref.~\cite{marcucci2016})
\begin{equation}\label{semigroup-duality}
\begin{aligned}
  \langle U_I(t)\varphi^{\pm} \mid \Psi_E^{\pm} \rangle &= \langle \varphi^{\pm} \mid (U_I(t))^\times \Psi_E^{\pm} \rangle\\& = e^{\pm iEt}\langle \varphi^{\pm} \mid \Psi_E^{\pm} \rangle \,\in\, \mathscr{H}_{\pm}^{2} \,\, \Longleftrightarrow\,\, \pm t \geq 0,
\end{aligned}
\end{equation}
where $\Psi_E^+ \equiv \psi_{\pm}^E$ and $\Psi_E^- \equiv \overline{\psi_{\pm}^E}$. In other words, analyticity within the respective Hardy spaces $\mathscr{H}_{\pm}^{2}$ demands that the full unitary operator splits into forward and backward temporal semigroups, $U_{I\pm}(t) = U_I(t)\big|_{\mathrm{S}_{\pm}}$, bounded to the future ($t \in \mathbb{R}_+$) and past ($t \in \mathbb{R}_-$), respectively.

Inheriting this structure, the total regularized evolution operator
\begin{equation}
    U_{\pm}(t) = \exp\left({-i\normalorder{H_b}t}\right)\big|_{\mathfrak{F}_b(\mathrm{S}_{\pm})},
\end{equation}
now dictates that the discrete excited states evolve as
\begin{equation}
  |n_{I}^{\pm}(t)\rangle = (U_{\pm}(t))^{\times}|n_{I}^{\pm}\rangle = e^{\mp n\Omega_{n\ell} t}|n_{I}^{\pm}\rangle,\quad \pm t\geq 0.
\end{equation}
In the Heisenberg picture, this translates directly to the generalized ladder operators via the commutator $[H_b, b^{\pm}_I] = \pm i \Omega_I b_I^{\pm}$, yielding\footnote{Without loss of generality, we set the initial condition $t_0=0$. For a realistic system in which the star is dynamically formed, $t_0$ should be the time at which $\xi \mathcal{R}$ becomes relevant to yield the tachyonic sector.}
\begin{equation}
    b_I^{\pm}(t) = b^{\pm}_I \exp(\mp \Omega_I t),\quad \pm t \geq 0.
\end{equation}
Consequently, at the quantum level, the field expansion in Eq.~\eqref{phi-b-eq} remains formally unchanged, but its constituent mode functions $v^{(\pm)}_{n\ell m}(x)$ are dynamically restricted to the past/future temporal domains $\pm t \geq 0$.

Crucially, this bounded semigroup dynamics ensures that the quantized field modes evade the pathological, exponential classical divergences as $|t| \to \infty$. Instead, selecting $\mathrm{S}_{+}$ physically isolates the decaying branch for forward evolution ($t \geq 0$), which spontaneously breaks time-reflection symmetry since a time-reversal transformation ($t \to -t$) maps the system out of its original functional space ($|n_I^{+}\rangle \to |n_I^{-}\rangle$).

With the total vacuum states $|{\bf 0}^{\pm}\rangle$ defined across both sectors by
\begin{equation}
    a_{\omega \ell m} (b^{\mp}_{n\bar{\ell} \bar{m}})^{\times}|{\bf 0}^{\pm}\rangle=0,\quad \forall \omega,n,\ell,m,\bar{\ell},\bar{m},
\end{equation}
the full vacuum fluctuations decouple as
\begin{equation}
    \langle {\bf 0}^{\mp}|\phi^2|{\bf 0}^{\pm}\rangle = \langle 0|\phi_a^2|0\rangle\underbrace{\langle 0^{\mp}|0^{\pm}\rangle}_{1} + \underbrace{\langle 0|0\rangle}_{1}\langle 0^{\mp}|\phi_b^2|0^{\pm}\rangle.
\end{equation}
While the classical contribution $\phi_b^2$ blows up exponentially, the quantum semi-group restriction yields the well-behaved distributional expectation value:
\begin{equation}\label{cured-fluctuation}
  \langle 0^{\mp}|\phi_b^2|0^{\pm}\rangle = \mp i\sum_{n\ell m} e^{\mp 2\Omega_{n\ell} t}|Y_{\ell m}(\theta,\varphi)|^2 r^{-2}\psi_{n \ell}^2(r), 
\end{equation}
for $\pm t\geq 0$. The restriction to these temporal semigroups eliminates the classical divergence, forcing the quantum fluctuations to decay exponentially as $t \to \pm\infty$. Furthermore, the purely imaginary nature of Eq.~\eqref{cured-fluctuation} reflects the fundamentally non-stationary and transient character of the tachyonic sector. Rather than representing a standard static observable, this complex fluctuation governs the dissipative structure of the vacuum, a feature that dynamically imprints a finite, real transition rate onto localized interacting probes, as we demonstrate in the next section

\section{Operational detection of the tachyonic sector}\label{sec:detector}

Since local vacuum expectation values such as $\langle 0^{\mp}|\phi_b^2|0^{\pm}\rangle$ yield purely imaginary results, they cannot be interpreted as direct physical measurements. To establish an operational interpretation, we implement an Unruh-DeWitt (UDW) particle detector framework to bridge the gap between the abstract rigged quantization and observable physical signatures~\cite{felipe2025udw}.

Let us model the probe as an idealized point-like two-level system moving along a classical worldline $x(\tau)=(t(\tau),{\bf x}(\tau))$, parameterized by its proper time $\tau$. The internal monopole moment operator evolves in the Heisenberg picture as $\mu(\tau)=|e\rangle\langle g|e^{i\Delta E \tau}+|g\rangle\langle e|e^{-i\Delta E \tau}$, where $\Delta E=E-E_0$ is the detector's energy gap between its ground state $|g\rangle$ and excited state $|e\rangle$. The interaction between the detector and the complete scalar field configuration $\phi = \phi_a + \phi_b$ is governed by the localized interaction Hamiltonian
\begin{equation}\label{interaction-hamiltonian}
  H_{\text{int}}(\tau) = \lambda\,\mu(\tau) \left[ \phi_a(x(\tau)) + \phi_b(x(\tau)) \right],
\end{equation}
where $\lambda \ll 1$ is a small coupling constant. The total kinematic state of the combined system resides within the composite space $\mathcal{H}_D \otimes \mathfrak{F}_a(\mathcal{H}) \otimes \mathfrak{F}_b^{\times}(\mathrm{S}^{\times})$, where the detector's Hilbert space $\mathcal{H}_D \cong \mathbb{C}^2$ is spanned by the orthogonal basis $\{|g\rangle,|e\rangle\}$. 

Initializing the system in its lowest-energy configuration $|g,{\bf 0}^{\pm}\rangle$, the selection of the tachyonic vacuum state (i.e., $\pm$ signs) is uniquely determined by the temporal domain of the trajectory $x(\tau)$. Specifically, when the proper time maps to the past coordinate domain ($t(\tau)<0$), the relevant initial state is $|{\bf 0}^{-}\rangle$, whereas when it maps to the future domain ($t(\tau)>0$), the relevant initial state is $|{\bf 0}^{+}\rangle$. To leading order in perturbation theory, the transition amplitude to a first excited state $|e,1_{\omega\ell m},1^{\pm}_{n\ell m}\rangle$ reads
\begin{equation}\label{amplitude}
    \mathcal{A}=\mathcal{A}_{a}+\mathcal{A}^{\pm},
\end{equation} 
where
\begin{equation}\label{amplitude-a}
    \mathcal{A}_{a}=-i\lambda\int_{-\infty}^{\infty} d\tau e^{i\Delta E \tau}\langle 1_{\omega\ell m}|\phi_a(x(\tau))|0\rangle\langle 1^{\mp}_{n\ell m}|0^{\pm}\rangle,
\end{equation}
and 
\begin{equation}\label{amplitude+-}
    \mathcal{A}^{\pm}=-i\lambda \int_{-\infty}^{\infty} d\tau e^{i\Delta E \tau}\langle 1_{\omega\ell m}|0\rangle\langle1^{\mp}_{n\ell m}|\phi_b^{\times}(x(\tau))|0^{\pm}\rangle.
\end{equation}

Due to the strict orthonormality between the oscillatory and tachyonic field sectors, i.e., $\langle1_{\omega \ell m} \mid 0\rangle = \langle 1_{n \ell m}^{\mp} \mid 0^{\pm}\rangle=0$, the interference terms vanish identically. Therefore, for a single first excited state, the absorption or emission of an energy quantum $\Delta E$ is a mutually exclusive process between $\phi_a$ and $\phi_b$. Furthermore, the generalized matrix elements $\langle 1^{\mp}_{n\ell m}|\phi_b|0^{\pm}\rangle$ inherently project the integration in Eq.~\eqref{amplitude+-} onto the restricted temporal domains of the semigroup modes,
\begin{equation}\label{time-selection}
      \langle 1_{n \ell m}^{\mp}|\phi_b^{\times}(x(\tau))|0^{\pm}\rangle = v^{(\pm)}_{n\ell m}(x(\tau)) \;\Theta(\pm t),   
\end{equation}
where $\Theta$ is the Heaviside step function. 

By summing over all final field states, the total transition probability splits as $\mathcal{P} = \mathcal{P}_a + \mathcal{P}^{\pm}$. Here, $\mathcal{P}_a$ represents the standard detector probability associated with the Wightman function $W_a(x,x^{\prime}) = \langle0|\phi_a(x)\phi_a(x^{\prime})|0\rangle$, while the novel tachyonic contribution is given by
\begin{equation}\label{probability+-}
  \mathcal{P}^{\pm} = \lambda^2 \int_{I_{\pm}} d\tau \int_{I_{\pm}} d\tau^{\prime}e^{-i\Delta E(\tau-\tau^{\prime})} W^{\pm}(x,x^{\prime}),
\end{equation}
with 
\begin{equation}\label{Wightman+-}
W^{\pm}(x,x^{\prime}) = \sum_{n\ell m} v^{(\pm)}_{n\ell m}(x(\tau)) v_{n\ell m}^{(\pm)\,\ast}(x(\tau^{\prime})).
\end{equation}
The integration domains are naturally bounded by the semigroups to $I_{+} := \{\tau \in \mathbb{R} \mid t(\tau) \geq 0\}$ and $I_{-} := \{\tau \in \mathbb{R} \mid t(\tau) \leq 0\}$.

To evaluate the transition probability explicitly, we assume the detector moves in a stationary circular orbit within the equatorial plane ($\theta=\pi/2$) at a constant radius $r_0 > R_s$ with angular velocity $\widetilde{\omega}_0 = d\varphi/dt$. Under these conditions, the proper time $\tau$ of the detector is linearly related to the coordinate time $t$ by $t(\tau)=\gamma\tau$, where $\gamma=(1-r_0^2\widetilde{\omega}_0^2)^{-1/2}$ is the Lorentz factor. Accordingly, the worldline of the UDW detector, parameterized by its proper time, is given by
\begin{equation}\label{worldline}
    x^{\mu}(\tau)=\left(\gamma \tau, r_0,\frac{\pi}{2},\widetilde{\omega}_0\gamma \tau \right).
\end{equation}

Notably, for $\widetilde{\omega}_0=0$, the detector is static and $\mathcal{P}_a$ vanishes, since inertial observers do not detect particle creation in the Minkowski vacuum. On the other hand, for $\widetilde{\omega}_0\neq 0$, $\mathcal{P}_a$ gives rise to the well-known nonvanishing circular Unruh-like background~\cite{delorenci2000,zhang2020,biermann2020}. As our goal is to understand the detector response due to the tachyonic modes, we henceforth focus on the additional contribution $\mathcal{P}^{\pm}$.

Using~\eqref{worldline} in the mode expansion~\eqref{Wightman+-}, the generalized Wightman function takes the form
\begin{equation}\label{generalized-wightman}
    W^{\pm}=\sum_{n\ell m} \frac{e^{\mp \gamma\Omega_{n\ell} (\tau+\tau^{\prime})+i m\gamma \widetilde{\omega}_0(\tau-\tau^{\prime})}}{2\Omega_{n\ell}}N_{\ell m}^2 [P_{\ell}^{m}(0)]^2r_0^{-2}\psi_{n\ell}^{2}(r_0),
\end{equation}
where $N_{\ell m}$ denotes the spherical harmonic normalization constant, and $P_{\ell}^{m}(0)$ is the associated Legendre polynomial evaluated at the equatorial plane $\theta=\pi/2$. Inserting Eq.~\eqref{generalized-wightman} into Eq.~\eqref{probability+-}, symmetry under time reflection implies $\mathcal{P}^{+} = \mathcal{P}^{-}$. For an infinite interaction time, considering both the past and future evolution, the total combined tachyonic response $\mathcal{P}_b = \mathcal{P}^{-} + \mathcal{P}^{+} = 2\mathcal{P}^{\pm}$ is obtained. We find that only modes with even $\ell+m$ contribute, since $P_{\ell}^{m}(0)=0$ for odd $\ell+m$, leading to
\begin{widetext}
\begin{equation}\label{Pb1}
    \mathcal{P}_b(\Delta E)=\lambda^2 \sum_{\substack{n\ell m\\ \ell+m\ \mathrm{even}}} \frac{2\ell +1}{\Omega_{n\ell}\left[\left(\Delta E-\gamma m \widetilde{\omega}_0\right)^2+\left(\gamma \Omega_{n\ell}\right)^2\right]}\frac{(\ell -m)!(\ell +m)!}{2^{2\ell}(4\pi)\left[\left(\frac{\ell-m}{2}\right)!\left(\frac{\ell+m}{2}\right)!\right]^2}r_0^{-2}\psi_{n\ell}^{2}(r_0).
\end{equation}
\end{widetext}

Equation \eqref{Pb1} reveals a rich spectroscopic structure: each tachyonic mode imprints an unnormalized Breit-Wigner (Lorentzian) distribution onto the detector's response. The resonance peaks are centered at the Doppler-shifted energies $E_R = \gamma m \widetilde{\omega}_0$, while the resonance widths are governed by the time-dilated instability rates, $\Gamma = 2\gamma \Omega_{n\ell}$. The UDW detector therefore operationally perceives each tachyonic mode as a physical resonant state. For a static probe ($\widetilde{\omega}_0=0, \gamma=1$), while the oscillatory contribution vanishes, $\mathcal{P}_a=0$, the tachyonic contribution remains nonzero, with a distribution centered at $\Delta E = 0$ with the intrinsic width $\Gamma = 2\Omega_{n\ell}$. 

In the language of resonance theory, the width dictates a characteristic transient lifetime $\tau_{\text{mean}} \sim 1/(2\gamma\Omega_{n\ell})$, meaning that the stable limit ($\Omega_{n\ell} \to 0$) smoothly restores a divergent lifetime. Furthermore, because the radial wavefunctions $\psi_{n\ell}(r_0)$ decay exponentially outside the stellar surface, the signal is strongly localized, requiring a close-in orbit ($r_0 \sim R_s$) for detection.

\begin{table}[h]
\centering
\renewcommand{\arraystretch}{1.2}
\setlength{\tabcolsep}{18pt}

\begin{tabular}{c|cc}
\hline
$\ell$ & $\Omega_{n\ell}$ & $C_{n\ell}$ \\
\hline
\multirow{2}{*}{0}
  & 0.875 & 0.475 \\
  & 4.237 & 3.387 \\
\hline
1 & 3.344 & 4.997 \\
\hline
2 & 1.763 & 1.362 \\
\hline
\end{tabular}
\caption{Dimensionless values of $\Omega_{n\ell}$ and $C_{n\ell}$ obtained from Eqs.~\eqref{matching2-eq} and~\eqref{normalization-psi}, respectively, for $\xi=10\xi_c=-10\pi^2/4$ and $R_s=1$.}
\label{tab:omega_c}
\end{table}

\begin{figure}[!htb]
    \centering
    \includegraphics[width=1\linewidth]{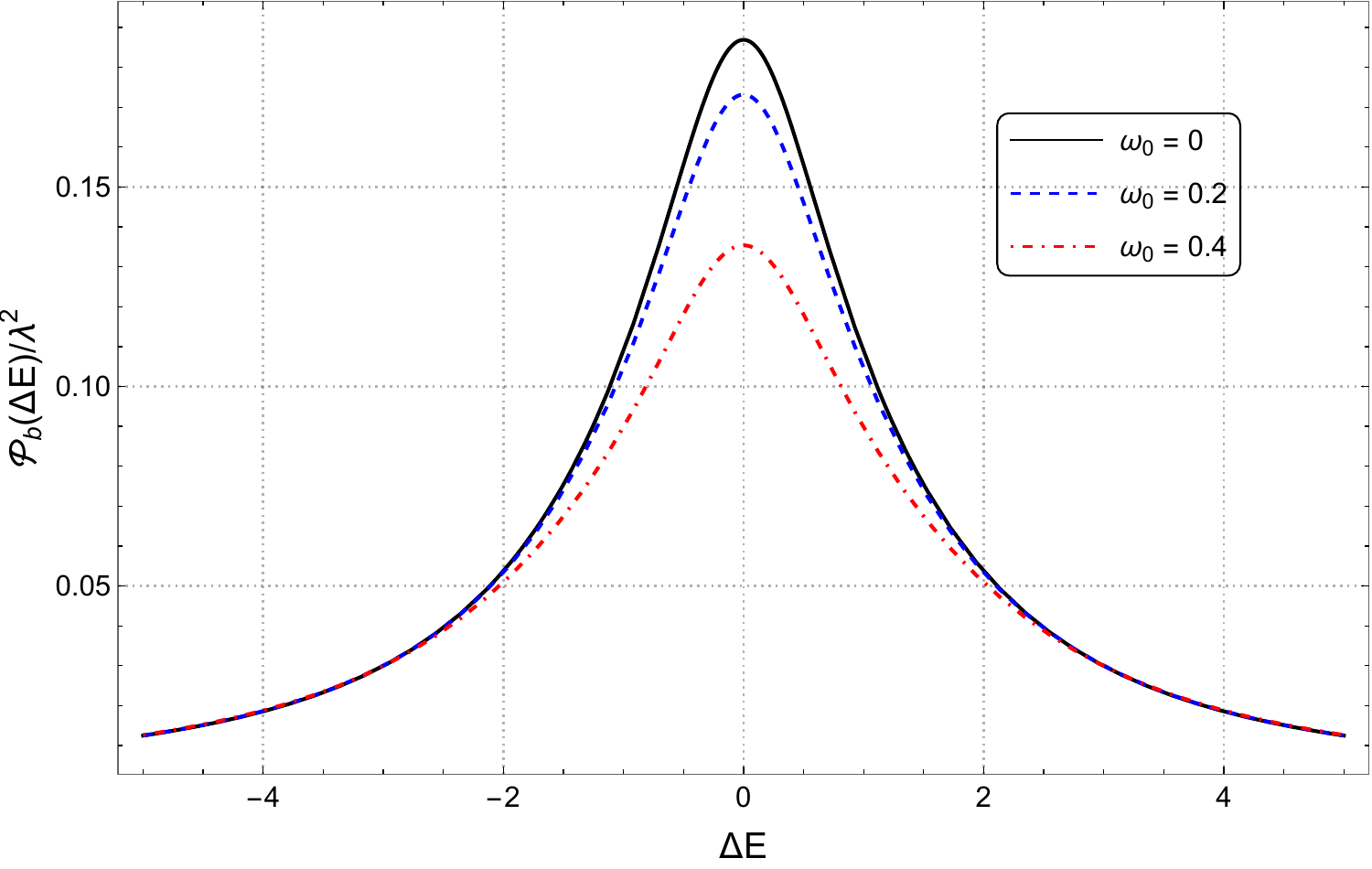}
    \caption{Tachyonic contribution to the detector response, $\mathcal{P}_b/\lambda^2$, as a function of $\Delta E$. The curves are computed from Eq.~\eqref{Pb1} using the characteristic values listed in Table~\ref{tab:omega_c} and $r_0=1.2R_s$. The solid, dashed, and dot-dashed curves correspond to orbital velocities $v=\widetilde{\omega}_0r_0\approx 0$, $0.2$, and $0.5$, respectively.}
    \label{fig:placeholder}
\end{figure}

To illustrate the total spectroscopic profile, we evaluate Eq.~\eqref{Pb1} numerically. For a coupling of $\xi=10\xi_c$, the non-vanishing tachyonic contributions are restricted to $\ell = 0, 1, 2$, as detailed in Table~\ref{tab:omega_c}. The geometric selection rule ($\ell+m$ even) isolates the active channels to $m=0$ for $\ell=0$; $m=\pm 1$ for $\ell =1$; and $m=\{0,\pm2\}$ for $\ell =2$, corresponding to resonant peaks at $E_R=0, \,\pm \gamma\widetilde{\omega}_0$, and $\pm 2\gamma\widetilde{\omega}_0$. Therefore, for $\ell\neq 0$, peaks appear in symmetric pairs around the origin. The resulting total response $\mathcal{P}_b/\lambda^2$ is plotted in Fig.~\ref{fig:placeholder} as a function of $\Delta E$ for a selection of angular velocities $\widetilde{\omega}_0=0, \,0.2,\, 0.4$ at $r_0 = 1.2R_s$. As the speed increases, the rotational motion lifts the resonance degeneracy, causing a distinct broadening and a corresponding drop in peak intensities. 

Crucially, the detector registers a strictly finite transition probability rather than a nonphysical, divergent particle production, even in the asymptotic limit of infinite interaction time. In this way, the RHS quantization successfully regularizes the unstable field sector, transforming the classical runaway catastrophe into a well-behaved, finite superposition of Breit-Wigner resonance channels that characterize the transient vacuum structure through a Lorentzian signature.

\section{Conclusions}\label{sec:conclusion}

In this work, we have discussed the quantization of tachyonic modes arising from a nonminimal coupling between a real scalar field and a constant-density star in the Newtonian limit. By mapping these modes onto a collection of decoupled inverted harmonic oscillators, we built a consistent Fock representation considering generalized quantum states within the dual spaces $\mathrm{S}_{\pm}^{\times}$ defined through the Gelfand triplets $\mathrm{S}_{\pm}\subset\mathcal{H}\subset\mathrm{S}_{\pm}^{\times}$. In this way, the corresponding quantum states are given by cross-orthonormalized resonant states with a structure that mirrors the classical tachyonic normalization according to the Klein-Gordon inner product and carries purely imaginary energies.

The primary impact of our approach lies in the quantum regularization of the classical runaway catastrophe, thereby avoiding the tachyonic instability. At the quantum level, the unitary time-evolution naturally splits into two distinct forward and backward temporal semigroups when constrained to the RHS. This mechanism confines the growing and decaying field branches to directed temporal domains, $\pm t\geq 0$. Consequently, vacuum fluctuations no longer exhibit exponential divergences as $|t|\to\infty$. Instead, the RHS framework forces the quantum fluctuations to decay exponentially, breaking time-reflection symmetry at the quantum level even within a stationary macroscopic background.

This regularized vacuum structure imprints a distinct signature when probed by a localized interacting system. By analyzing a circular orbiting UDW detector, we showed that the detector perceives each discrete tachyonic model not as an infinite particle-production bath, but as a physical resonant system. Hence, the resulting response function is a finite superposition of Lorentzian distributions in which the rotational motion of the probe induces a resonance degeneracy, inducing a Doppler shift that centers the peaks at $E_R=\gamma m \widetilde{\omega}_0$ while the time-dilated tachyonic energies dictate the resonance widths, $\Gamma=2\gamma\Omega_{n\ell}$. This Lorentzian signature profile provides a clear, finite, and identifiable blueprint for detecting tachyonic modes.

Looking forward, our framework opens several promising avenues for future research. First, while our weak-field stellar model provides a clean background in which we can isolate the core physics of the RHS quantization, a natural next step is to apply this machinery to fully relativistic compact objects, such as neutron stars near the threshold of spontaneous scalarization. In such scenarios, the strong background curvature and the dynamical formation of the star could map directly onto the activation time $t_0$ governing the semigroup dynamics. Second, our results hold significant potential for exploration within analog gravity systems, such as acoustic black holes in Bose-Einstein condensates or traveling-wave structures in metamaterials~\cite{barcelo2011,almeida2023,ribeiro2020,syu2024}. Probing these analog systems with localized detectors could provide a way to observe the predicted Breit-Wigner response, connecting abstract quantum field theory in curved spacetimes with laboratory experiments.

\section*{Data availability}
This is a purely mathematical work and no data was created or analyzed in this study. All figures can be reproduced directly from the presented equations.

\acknowledgments
B. S. F. acknowledges support from the São Paulo Research Foundation (FAPESP, Brazil), Grant 2025/19841-2. M. R. acknowledges partial support from the Conselho Nacional de Desenvolvimento Científico e Tecnológico (CNPq, Brazil), Grant 315991/2023-2, and from the São Paulo Research Foundation (FAPESP, Brazil), Grant 2024/00923-6.

\appendix

\section{Effective Lagrangian of the tachyonic sector}
\label{app:effective-lagrangian}

To construct the effective Lagrangian for the tachyonic sector, consider the mode \eqref{alternative-modes} into the Lagrangian density \eqref{lagrangian-eq} and integrate over the spatial volume. In this case, the metric components are simply $g^{\mu\nu} \approx \text{diag}(-1, 1, r^{-2}, r^{-2}\sin^{-2}\theta)$, with $\sqrt{-g} \approx r^2 \sin\theta$, allowing the action to be recast as $S_I = \int dt \, L_I(t)$, where:
\begin{equation}\label{L_full_app}
  \begin{aligned}
      L_I(t) = \frac{1}{2} \int_0^\infty \! dr \int \! d\Omega \, r^2 \bigg[ (\partial_t \phi_I)^2 - (\partial_r \phi_I)^2 \\- \frac{1}{r^2}(\nabla_{\Omega}\phi_I)^2 - \xi R \phi_I^2 \bigg].
  \end{aligned}
\end{equation}
Using the orthonormality of the real spherical harmonics $Z_{\ell m}$, i.e., $\int d\Omega \, Z^2_{\ell m} = 1$, along with the angular gradient identity $\int d\Omega (\nabla_\Omega Z_{\ell m})^2 = \ell(\ell+1)$, the angular integration collapses Eq.~\eqref{L_full_app} into a strictly radial representation:
\begin{equation}\label{L_radial_app}
\begin{aligned}
  L_I(t) = \frac{\dot{\chi}_I^2}{2} \int_0^\infty \! \psi^2 \, dr - \frac{\chi_I^2}{2} \int_0^\infty \! dr \bigg[ r^2 \left( \frac{\psi^{\prime}}{r} - \frac{\psi}{r^2} \right)^2\\ + \frac{\ell(\ell+1)}{r^2}\psi^2 + \xi R \psi^2 \bigg],
\end{aligned}
\end{equation}
where the first integral equals unity by virtue of the radial normalization condition \eqref{normalization-psi}.

The second integration in Eq.~\eqref{L_radial_app} can be evaluated by expanding the quadratic term and performing a standard integration by parts under regular boundary conditions ($\psi(0) = \psi(\infty) = 0$). This procedure simplifies the integrand via the identity
\begin{equation}
    \int_0^\infty r^2 \left( \frac{\psi^{\prime}}{r} - \frac{\psi}{r^2} \right)^2 dr = -\int_0^\infty \psi \psi^{\prime\prime} dr,
\end{equation}
yielding the total Lagrangian
\begin{equation}
  L_I(t) = \frac{\dot{\chi}_I^2}{2} - \frac{\chi_I^2}{2} \int_0^\infty dr \, \psi \left[ -\psi^{\prime\prime} + \left( \frac{\ell(\ell+1)}{r^2} + \xi R \right) \psi \right].
\end{equation}
Finally, using the radial eigenvalue equation \eqref{radial-eq}, with $\omega^2=-\Omega^2$, the remaining integral reduces to the effective single-mode tachyonic Lagrangian:
\begin{equation}
  L_I(t) = \frac{1}{2} \left( \dot{\chi}_I^2 + \Omega_{I}^2 \chi_I^2 \right).
\end{equation}
Summing over all modes, we finally obtain Eq.~\eqref{lagrangian-phib}.

\end{document}